\documentclass[aps,prl,twocolumn,showpacs,amsmath,preprintnumbers,amssymb,groupedaddress,superscriptaddress]{revtex4-1}
\usepackage{graphicx}
\usepackage{dcolumn}
\usepackage{amsmath}

\usepackage{hyperref}
\usepackage{breakurl}

\def\pT{\mbox{$p_{\rm T}$ }}  

\def\GeVc{\mbox{GeV/$c$}}
\def\sNN{\mbox{$\sqrt{s_{_{\rm NN}}}$ }}
\newcommand{ \be }{\begin{equation}}      
\newcommand{ \ee }{\end{equatiton}}      
\newcommand{ \bea }{\begin{eqnarray}}      
\newcommand{ \eea }{\end{eqnarray}}

\usepackage{color}

\begin{document}


\title{
Measurement of $D^0$ azimuthal anisotropy at mid-rapidity in Au+Au collisions at \sNN = 200\,GeV}




\affiliation{AGH University of Science and Technology, FPACS, Cracow 30-059, Poland}
\affiliation{Argonne National Laboratory, Argonne, Illinois 60439}
\affiliation{Brookhaven National Laboratory, Upton, New York 11973}
\affiliation{University of California, Berkeley, California 94720}
\affiliation{University of California, Davis, California 95616}
\affiliation{University of California, Los Angeles, California 90095}
\affiliation{Central China Normal University, Wuhan, Hubei 430079}
\affiliation{University of Illinois at Chicago, Chicago, Illinois 60607}
\affiliation{Creighton University, Omaha, Nebraska 68178}
\affiliation{Czech Technical University in Prague, FNSPE, Prague, 115 19, Czech Republic}
\affiliation{Nuclear Physics Institute AS CR, 250 68 Prague, Czech Republic}
\affiliation{Frankfurt Institute for Advanced Studies FIAS, Frankfurt 60438, Germany}
\affiliation{Institute of Physics, Bhubaneswar 751005, India}
\affiliation{Indiana University, Bloomington, Indiana 47408}
\affiliation{Alikhanov Institute for Theoretical and Experimental Physics, Moscow 117218, Russia}
\affiliation{University of Jammu, Jammu 180001, India}
\affiliation{Joint Institute for Nuclear Research, Dubna, 141 980, Russia}
\affiliation{Kent State University, Kent, Ohio 44242}
\affiliation{University of Kentucky, Lexington, Kentucky, 40506-0055}
\affiliation{Lamar University, Physics Department, Beaumont, Texas 77710}
\affiliation{Institute of Modern Physics, Chinese Academy of Sciences, Lanzhou, Gansu 730000}
\affiliation{Lawrence Berkeley National Laboratory, Berkeley, California 94720}
\affiliation{Lehigh University, Bethlehem, PA, 18015}
\affiliation{Max-Planck-Institut fur Physik, Munich 80805, Germany}
\affiliation{Michigan State University, East Lansing, Michigan 48824}
\affiliation{National Research Nuclear University MEPhI, Moscow 115409, Russia}
\affiliation{National Institute of Science Education and Research, HBNI, Jatni 752050, India}
\affiliation{National Cheng Kung University, Tainan 70101 }
\affiliation{Ohio State University, Columbus, Ohio 43210}
\affiliation{Institute of Nuclear Physics PAN, Cracow 31-342, Poland}
\affiliation{Panjab University, Chandigarh 160014, India}
\affiliation{Pennsylvania State University, University Park, Pennsylvania 16802}
\affiliation{Institute of High Energy Physics, Protvino 142281, Russia}
\affiliation{Purdue University, West Lafayette, Indiana 47907}
\affiliation{Pusan National University, Pusan 46241, Korea}
\affiliation{Rice University, Houston, Texas 77251}
\affiliation{University of Science and Technology of China, Hefei, Anhui 230026}
\affiliation{Shandong University, Jinan, Shandong 250100}
\affiliation{Shanghai Institute of Applied Physics, Chinese Academy of Sciences, Shanghai 201800}
\affiliation{State University Of New York, Stony Brook, NY 11794}
\affiliation{Temple University, Philadelphia, Pennsylvania 19122}
\affiliation{Texas A\&M University, College Station, Texas 77843}
\affiliation{University of Texas, Austin, Texas 78712}
\affiliation{University of Houston, Houston, Texas 77204}
\affiliation{Tsinghua University, Beijing 100084}
\affiliation{University of Tsukuba, Tsukuba, Ibaraki, Japan,}
\affiliation{Southern Connecticut State University, New Haven, CT, 06515}
\affiliation{United States Naval Academy, Annapolis, Maryland, 21402}
\affiliation{Valparaiso University, Valparaiso, Indiana 46383}
\affiliation{Variable Energy Cyclotron Centre, Kolkata 700064, India}
\affiliation{Warsaw University of Technology, Warsaw 00-661, Poland}
\affiliation{Wayne State University, Detroit, Michigan 48201}
\affiliation{World Laboratory for Cosmology and Particle Physics (WLCAPP), Cairo 11571, Egypt}
\affiliation{Yale University, New Haven, Connecticut 06520}

\author{L.~Adamczyk}\affiliation{AGH University of Science and Technology, FPACS, Cracow 30-059, Poland}
\author{J.~K.~Adkins}\affiliation{University of Kentucky, Lexington, Kentucky, 40506-0055}
\author{G.~Agakishiev}\affiliation{Joint Institute for Nuclear Research, Dubna, 141 980, Russia}
\author{M.~M.~Aggarwal}\affiliation{Panjab University, Chandigarh 160014, India}
\author{Z.~Ahammed}\affiliation{Variable Energy Cyclotron Centre, Kolkata 700064, India}
\author{N.~N.~Ajitanand}\affiliation{State University Of New York, Stony Brook, NY 11794}
\author{I.~Alekseev}\affiliation{Alikhanov Institute for Theoretical and Experimental Physics, Moscow 117218, Russia}\affiliation{National Research Nuclear University MEPhI, Moscow 115409, Russia}
\author{D.~M.~Anderson}\affiliation{Texas A\&M University, College Station, Texas 77843}
\author{R.~Aoyama}\affiliation{University of Tsukuba, Tsukuba, Ibaraki, Japan,}
\author{A.~Aparin}\affiliation{Joint Institute for Nuclear Research, Dubna, 141 980, Russia}
% ...

\collaboration{STAR Collaboration}\noaffiliation


\date{\today}

\begin{abstract}
We report the first measurement of the elliptic anisotropy ($v_2$) of the charm meson $D^0$ at mid-rapidity ($|y|$\,$<$\,1) in Au+Au collisions at \sNN = 200\,GeV. The measurement was conducted by the STAR experiment at RHIC utilizing a new high-resolution silicon tracker. The measured $D^0$ $v_2$ in 0--80\% centrality Au+Au collisions can be described by a viscous hydrodynamic calculation for transverse momentum ($p_{\rm T}$) less than 4\,GeV/$c$. The $D^0$ $v_2$ as a function of transverse kinetic energy ($m_{\rm T} - m_0$, where $m_{\rm T} = \sqrt{p_{\rm T}^2 + m_0^2}$) is consistent with that of light mesons in 10--40\% centrality Au+Au collisions. 
These results suggest that charm quarks have achieved local thermal equilibrium with the medium created in such collisions. Several theoretical models, with the temperature--dependent, dimensionless charm spatial diffusion coefficient ($2{\pi}TD_s$) in the range of $\sim$2--12, are able to simultaneously reproduce our $D^0$ $v_2$ result and our previously published results for the $D^0$ nuclear modification factor.
\end{abstract}


\maketitle



Quantum chromodynamics (QCD) is a non-Abelian gauge theory which describes the strong interactions between quarks and gluons. Experiments at the Relativistic Heavy Ion Collider (RHIC) and the Large Hadron Collider (LHC) indicate that a novel form of QCD matter, consistent with a strongly coupled Quark-Gluon Plasma (sQGP), is created in heavy-ion collisions at these energies~\cite{StarWhitePaper,PhenixWhitePaper,LhcSummary}. A key piece of evidence for this new state of matter is the strong collective, anisotropic flow of produced light flavor particles, suggesting possibly hydrodynamic behavior of the strongly interacting matter during the collision~\cite{Gale:2012rq}. 


Heavy quarks (charm and bottom) are predominantly created in the initial hard scatterings in a heavy-ion collision, and 
their propagation in the sQGP can be described as Brownian-like motion~\cite{Moore:2004tg,Rapp:2008qc}. 
The sQGP properties can be accessed through experimental observables such as the nuclear modification factor ($R_{\rm AA}$)~\cite{Wang:1991xy}, the ratio of the yield in heavy-ion collisions to the scaled yield in proton+proton ($p$+$p$) collisions, and the elliptic anisotropy ($v_2$)~\cite{Poskanzer98}, the second Fourier coefficient of the particle yield with respect to the reaction plane (defined by the beam axis and the direction of the impact parameter between two colliding nuclei). 
Of these observables, the $v_2$ at low transverse momentum ($p_{\rm T}$) where light and strange flavor hadrons appear to behave hydrodynamically, is of particular interest because it probes the properties of the bulk medium in the strongly-coupled region and is less affected by the shadowing and Cronin effects~\cite{HFSummary}.

Recent measurements at RHIC and the LHC show that high-\pT charm hadron yields are significantly suppressed in central heavy-ion collisions indicating strong charm--medium interactions~\cite{StarRaa,AliceRaa,AliceRaa2}. 
The $D$-meson $v_2$ measured by ALICE~\cite{AliceV2,*AliceV22} is comparable to that of light hadrons at the LHC.
So far, charm quark flow at RHIC has only been inferred from measurements of semileptonic decays of charm and bottom hadrons~\cite{PhenixNpe,*PhenixNpe2,STARNpe}. However, a clear interpretation of lepton $v_2$ measurements suffers from an ambiguity in the lepton sources between charm and bottom decays and the decay kinematics. On the other hand, there has been significant progress in theoretical calculations for charm hadron $v_2$ in heavy-ion collisions~\cite{PHSD,*PHSD2,SUBATECH,*SUBATECH2,*SUBATECHPrivateCom,Torino,*Torino2,Duke,*DukePrivateCom,TAMU,*TAMU2,*TAMUPrivateCom,LBT,BAMPS,*BAMPS2,LANL}. A precise measurement of charm hadron $v_2$ over a wide momentum range is expected to provide valuable insights into the sQGP properties~\cite{HFSummary}.

In this Letter, we report the first measurement of the $D^0$ anisotropy parameter $v_2$ at mid-rapidity ($|y|$\,$<$\,1) at RHIC by the STAR Collaboration using the newly completed Heavy Flavor Tracker (HFT)~\cite{HFT,HFTQM14}. The HFT is a high-resolution silicon detector system, which aims for the topological reconstruction of secondary decay vertices of open heavy flavor hadrons. It has three sub-detectors: the Silicon Strip Detector, the Intermediate Silicon Tracker (IST), and the Pixel (PXL) detector. 
In the 2014 Au+Au run at \sNN = 200\,GeV, $\sim$\,1.1\,billion minimum bias triggered events, selected by a coincidence signal between the east and west Vertex Position Detectors (VPD)~\cite{VPD} located at 4.4\,$<|\eta|<$\,4.9 ($\eta$ is the pseudo-rapidity), were recorded with the IST and the PXL.
In this analysis, the reconstructed collision primary vertex (PV) is required to be less than 6\,cm from the detector center along the beam axis to ensure good HFT acceptance. The collision centrality, the fraction of the total hadronic cross section, is defined using the measured charged track multiplicity at mid-rapidity and corrected for the online VPD triggering inefficiency using a Monte Carlo Glauber simulation~\cite{MCGlauber}.

$D^0$ and $\overline{D}^0$ mesons are reconstructed in the $K^{\mp}\pi^{\pm}$ decay channel, which has a short proper decay length ($c\tau$\,$\sim$\,123\,$\mu$m)~\cite{PDG}. Charged tracks are reconstructed by the Time Projection Chamber (TPC)~\cite{TPC} together with the HFT in a 0.5\,T uniform magnetic field. Tracks are required to have a minimum of 20 TPC hits (out of a maximum of 45), hits in all layers of PXL and IST sub-detectors, \pT $>$ 0.6\,GeV/$c$, and $|\eta|$\,$<$\,1. To identify particle species, the ionization energy loss, $dE/dx$, measured by the TPC is required to be within three and two standard deviations from the expected values for $\pi$ and $K$, respectively. The particle identification is extended by the Time Of Flight (TOF)~\cite{TOF} detector up to $p_{\rm T}\sim 1.6$\,GeV/$c$ by requiring the $1/\beta$ ($\beta$ is particle velocity in unit of the speed of light), calculated from the path length and the TOF, to be less than three standard deviations different from the expected value calculated using the $\pi$ or $K$ mass and the measured momentum.

\begin{figure}
\includegraphics[width=1.\columnwidth]{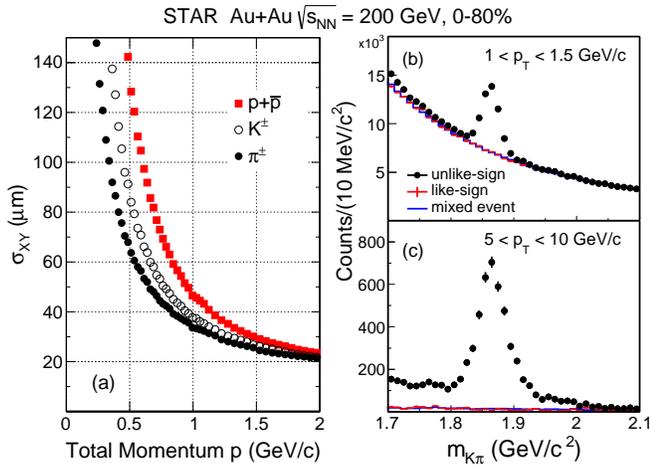}%
\caption{(color online) Identified particle pointing resolution in the transverse plane as a function of particle momentum (a). Invariant mass spectra of $K\pi$ pairs for 1\,$<$\,$p_{\rm T}$\,$<$\,1.5\,GeV/$c$ (b) and 5\,$<$\,$p_{\rm T}$\,$<$\,10\,GeV/$c$ (c), respectively. The solid data points are the $D^0$ signal reconstructed with unlike-sign pairs. The red crosses and the blue lines show the like-sign and mixed event background distributions. \label{fig:D0Signal}}
\end{figure}

Figure~\ref{fig:D0Signal} (a) shows the track pointing resolution to the collision vertex in the transverse plane ($\sigma_{\rm XY}$) as a function of momentum ($p$) for identified particles in 0--80\% centrality Au+Au collisions at \sNN = 200\,GeV. The resolution is better than 55\,$\mu$m for kaons with $p\geq$ 0.75\,GeV/$c$. 
With two daughter tracks, a secondary decay vertex can be reconstructed as the middle point on the distance of the closest approach (DCA) between them. 
The primary background is due to fake pairs coming from random combinations of tracks which propagate directly from the collision point.  The background can be significantly reduced by applying cuts on five variables:
decay length (the distance between the decay vertex and the PV), DCA between the two daughters, DCA between the reconstructed $D^0$ track and the PV, DCA between the $\pi$ track and the PV, and the DCA between the $K$ track and the PV. The cuts on these variables are optimized using the Toolkit for Multivariate Data Analysis (TMVA) package~\cite{TMVA}. Their optimization was pursued separately in each $D^0$ candidate \pT bin in order to have the greatest signal significance. 


Figures~\ref{fig:D0Signal} (b) and (c) show the invariant mass spectra of $K\pi$ pairs after applying these cuts for two \pT bins. Comparing these mass spectra with the previous $D^0$ study~\cite{StarRaa}, the signal significance is markedly improved due to the background rejection using the geometric cuts enabled by the HFT ($\sim$220$\sigma$ vs. $\sim$13$\sigma$ per billion events). The combinatorial background is estimated with like-sign $K\pi$ pairs and the mixed event unlike-sign technique in which $K$ and $\pi$ with opposite charge signs from different events are paired. The mixed event distributions are normalized to the like-sign distributions in the mass range of 1.7--2.1\,GeV/$c^2$. The remaining contributions to the background is expected to come from the correlated sources, e.g. $K\pi$ pairs from jet fragments or multi-prong decays of heavy flavor mesons.

\begin{figure}
\includegraphics[width=1.\columnwidth]{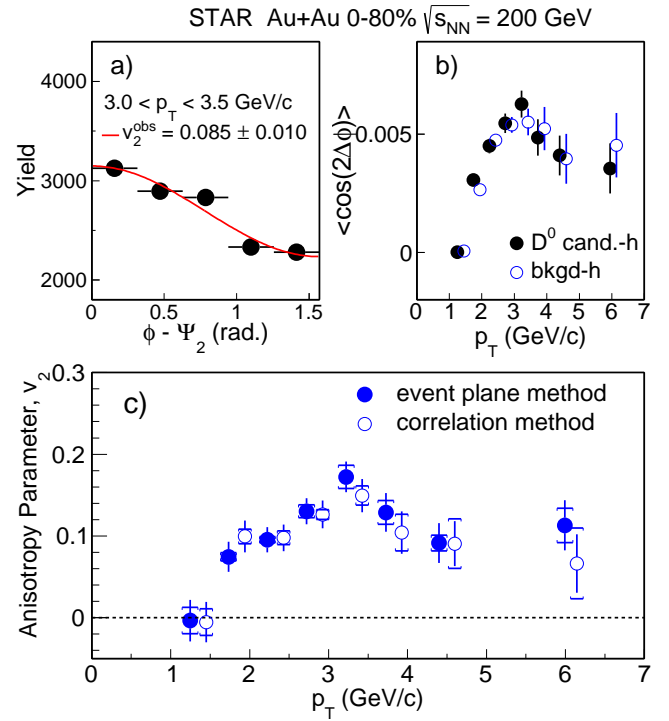}%
\caption{(color online) (a) $D^0$ yield as a function of $\phi-\Psi_2$ fit to $A(1+2v_2\mathrm{cos}(2(\phi-\Psi_2)))$, for 3\,$<$\,$p_{\rm T}$\,$<$\,3.5\,GeV/$c$. (b) Correlations ${\langle}\cos(2\Delta\phi)\rangle$ between the $D^0$ candidate or background and charged particles, as a function of $p_{\rm T}$. (c) $v_{2}$ as a function of \pT for $D^0$ calculated with the event plane and correlation methods. The data shown in all three panels are for 0--80\% centrality Au+Au collisions at \sNN = 200\,GeV. The vertical bars and the brackets represent the statistical and systematic uncertainties, respectively. The estimated non-flow contribution is not shown in this plot, but is common to both methods. In (a) and (b), only statistical uncertainties are shown as vertical bars (not visible if they are smaller than marker sizes). In (b) and (c), the open points are shifted along the $x$-axis for clarity.
\label{fig:v2Methods}}
\end{figure}

Two different methods are employed to calculate $v_2$: the event plane method~\cite{Poskanzer98} and the correlation method~\cite{correlationMethod,correlationMethod2}. In the event plane method, a second order event plane angle $\Psi_2$ is reconstructed from TPC tracks excluding decay products of $D^0$ mesons and after correcting for the azimuthal nonuniformity in the detector efficiency~\cite{Poskanzer98}.  To suppress non-flow effects (correlations not connected to the event plane, such as resonance decays and jet correlations), 
only particles from the opposite $\eta$ hemisphere of the reconstructed $D^0$ and outside of an additional $\eta$-gap of $|\Delta\eta|>$ 0.05 are used in the event plane reconstruction. 
The $D^0$ yields are measured in azimuthal bins relative to the event plane azimuth ($\phi-\Psi_2$). The yields are weighted by $1/(\varepsilon\times R)$, where $\varepsilon$ is the $D^0$ reconstruction efficiency$\times$acceptance and $R$ the event plane angle resolution~\cite{Poskanzer98} for each centrality interval~\cite{Masui:2012zh}. 
In each $\phi-\Psi_2$ bin, the mixed event background, scaled to the like-sign background, is subtracted from the unlike-sign distribution. The $D^0$ yield is obtained via the side band method by subtracting the scaled counts in two invariant mass ranges around the signal (1.71$-$1.80 and 1.93$-$2.02\,GeV/$c^2$) from the counts in the signal region (1.82$-$1.91\,GeV/$c^2$)~\cite{Adamczyk:2012af}. A fit method using a Gaussian function for $D^0$ signal plus a first order polynomial function for the background is also used to estimate the systematic uncertainty on the raw yield extraction. 
Figure~\ref{fig:v2Methods} (a) shows an example of the weighted $D^0$ yield as a function of $\phi-\Psi_2$. 
The observed $v_{2}$ is then obtained by fitting with a functional form $A(1+2v_2\mathrm{cos}(2(\phi-\Psi_2)))$, where $A$ is a normalization parameter. Finally, the true $v_2$ is obtained by scaling the observed $v_2$ with $\langle1/R\rangle$ to correct for the event plane angle resolution~\cite{Masui:2012zh}.

In the correlation method~\cite{correlationMethod,correlationMethod2}, $v_2$ is calculated for $D^0$ candidates and the background, separately. For example, the $D^0$ candidate-hadron azimuthal cumulant \mbox{$V_{2}^{\rm cand-h} \equiv {\langle}\cos(2\phi_{\rm cand}-2\phi_{h}){\rangle}$}, shown as a function of \pT as solid markers in Fig.~\ref{fig:v2Methods} (b), is calculated by the $Q$-cumulant method where $\phi_{\rm cand}$ and $\phi_{h}$ are azimuthal angles for $D^0$ candidates and charged hadrons, respectively~\cite{correlationMethod2}. The average is taken over all events and all particles. Neglecting non-flow contributions, the following factorization can be assumed to obtain the $D^0$ $v_2$: $V_{2}^{\rm cand-h} = v_{2}^{\rm cand}v_{2}^{h}$. Here, $v_{2}^{h}$ can be obtained from hadron-hadron correlations via $V_{2}^{h-h}=v_{2}^{h}v_{2}^{h}$. The same $\eta$-gap as in the event plane method was chosen for the correlation analysis.
The $D^0$ background $v_2$ is calculated similarly, with the background represented by the average of the like-sign $K\pi$ pairs in the $D^{0}$ mass window ($\pm3\sigma$, where $\sigma$ is the signal width) and side bands (4$-$9$\sigma$ away from the $D^0$ peak, both like-sign and unlike-sign $K\pi$ pairs). The background-hadron cumulant is also shown in Fig.~\ref{fig:v2Methods} (b) as open circles. The $D^0$ $v_2$ is obtained from the candidate and background $v_2$ and their respective yields ($N_{\rm cand}$, $N_{\rm bg}$) 
by \mbox{$v_{2}=(N_{\rm cand}v_{2}^{\rm cand}-N_{\rm bg}v_{2}^{\rm bg})/(N_{\rm cand}-N_{\rm bg})$}.




The systematic uncertainty is estimated by comparing $v_2$ obtained from the following different methods: a) the fit vs. side-band methods, b) varying invariant mass ranges for the fit and for the side bands, c) varying geometric cuts so that the efficiency changes by $\pm$50\% with respect to the nominal value. 
These three different sources are varied independently to form multiple combinations.
We then take the maximum difference from these combinations and divide by $\sqrt{12}$ as one standard deviation of the systematic uncertainty. 
The feed-down contribution from $B$-meson decays to our measured $D^0$ yield is estimated to be less than 4\%.
Compared to other systematic uncertainties, this contribution is negligible even in the extreme case that $B$-meson $v_2$ is 0.

Figure~\ref{fig:v2Methods} (c) shows the result of the $D^0$ $v_2$ in 0--80\% centrality Au+Au events as a function of $p_{\rm T}$. The results from the event plane and correlation methods are consistent with each other within uncertainties. For further discussion in this letter, we use $v_2$ from the event plane method only, which has been widely used in previous STAR identified particle $v_2$ measurements~\cite{ksV2,phiOmegaV2}.

\begin{figure}
\includegraphics[width=1.\columnwidth]{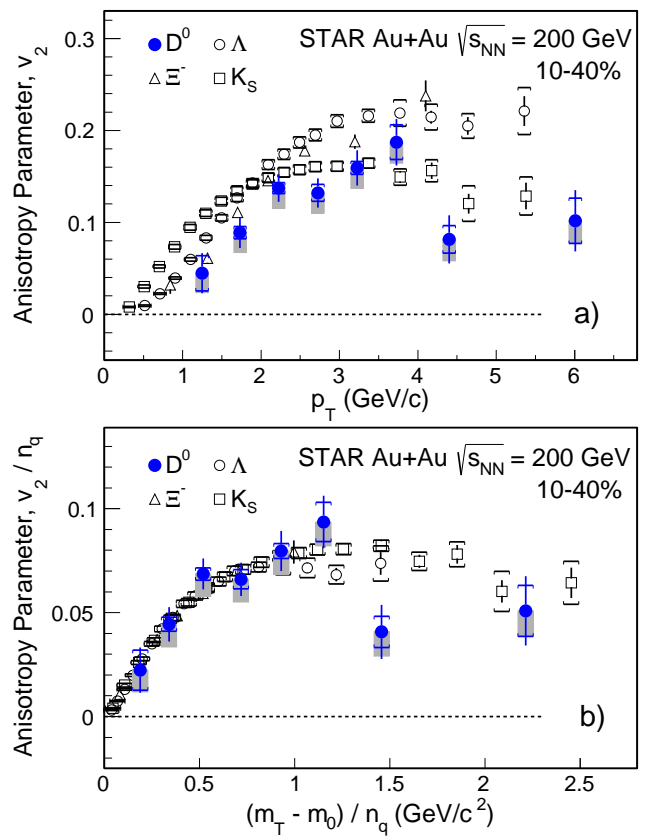}%
\caption{(color online) (a) $v_2$ as a function of \pT and (b) $v_2/n_q$ as a function of $(m_{\rm T}-m_0)/n_q$ for $D^0$ in 10--40$\%$ centrality Au+Au collisions compared with $K^0_S$, $\Lambda$, and $\Xi^-$~\cite{ksV2}. 
The vertical bars and brackets represent statistical and systematic uncertainties, and the grey bands represent the estimated non-flow contribution.
\label{fig:v2CompareWithData}}
\end{figure}

\begin{figure}
\includegraphics[width=1.\columnwidth]{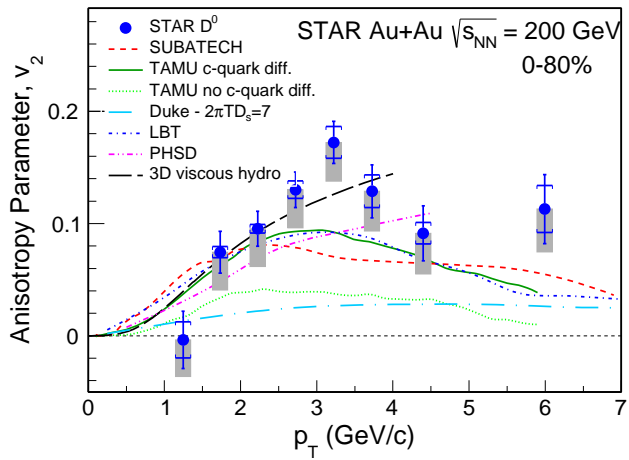}%
\caption{(color online) $v_{2}$ as a function of \pT for $D^0$ in 0--80$\%$ centrality Au+Au collisions compared with model calculations~\cite{PHSD,*PHSD2,SUBATECH,*SUBATECH2,*SUBATECHPrivateCom,Torino,*Torino2,hydroPang,*hydroPang2,*hydroPrivateCom,TAMU,*TAMU2,*TAMUPrivateCom,Duke,*DukePrivateCom,LBT}.
\label{fig:v2CompareWithModel}}
\end{figure}

The residual non-flow contribution is estimated by scaling the $D^0$-hadron correlation (with the same $\eta$ gap used in the analysis) in $p$+$p$ collisions, where only the non-flow effects are present, by the average $v_2$ ($\overline{v_2}$) and multiplicity ($M$) of charged hadrons used for event plane reconstruction or $D^0$-hadron correlations in Au+Au collisions. Thus the non-flow contribution is estimated to be \mbox{$\big\langle\displaystyle\sum\nolimits_{i}\cos2(\phi_{D^0}-\phi_i)\big\rangle/M\overline{v_2}$}~\cite{Adams:2004wz}, where $\phi_{D^0}$ and $\phi_i$ are the azimuthal angles for the $D^0$ and hadron, respectively. The $\sum\nolimits_{i}$ is done for charged tracks in the same event, and $\langle\rangle$ is an average over all events. The $D^0$-hadron correlation in $p$+$p$ collisions is deduced from $D^*$-hadron correlations measured with data taken by STAR in year 2012 for $p_{\rm T}$\,$>$\,3\,\GeVc~
and from a PYTHIA simulation for $p_{\rm T}$\,$<$\,3\,GeV/$c$. The correlations in $p$+$p$ collisions were used as a conservative estimate
since the correlation may be suppressed in Au+Au collisions due to the hot medium effect.
The estimated non-flow contribution is shown separately (grey bands) along with the systematic and statistical uncertainties in Figs.~\ref{fig:v2CompareWithData} and \ref{fig:v2CompareWithModel}.

For cross check we performed a MC simulation using the measured $D^0$ $v_2$ to calculate the single electron $v_2$ and compare to previous RHIC measurements~\cite{PhenixNpe,PhenixNpe2,STARNpe}. Both the PHENIX and STAR measurements are compatible with the calculated electron $v_2$ at $p_{\rm T}<$ 3\,GeV/$c$ where the charm hadron contribution dominates~\cite{FONLL,STAReh,PHENIXecb}. At higher $p_{\rm T}$ region, where the bottom contribution is sizable, the large uncertainty in the measurement of $v_2$ of single electrons does not allow for a reasonable extraction of $v_2$ for $B$-mesons. 

Figure~\ref{fig:v2CompareWithData} compares the measured $D^0$ $v_2$ from the event plane method in 10--40$\%$ centrality bin with $v_2$ of $K_{S}^0$, $\Lambda$, and $\Xi^-$~\cite{ksV2}. The comparison between $D^0$ and light hadrons needs to be done in a narrow centrality bin 
to avoid the bias caused by the fact that the $D^0$ yield scales with number of binary collisions while the yield of light hadrons scales approximately with number of the participants~\cite{Nasim:2016}.
Panel (a) shows $v_2$ as a function of \pT where a clear mass ordering for $p_{\rm T}$\,$<$\,2\,GeV/$c$ including $D^0$ mesons is observed. For $p_{\rm T}$\,$>$\,2\,GeV/$c$, the $D^0$ meson $v_2$ follows that of other light mesons indicating significant charm quark flow at RHIC~\cite{NCQscaling,ksV2,phiOmegaV2}. 
Recent ALICE measurements show that the $D^0$ $v_2$ is comparable to that of charged hadrons in 0-50\% Pb+Pb collisions at \sNN = 2.76\,TeV~\cite{AliceV2,AliceV22} suggesting sizable charm flow at the LHC.
Panel (b) shows $v_2/n_q$ as a function of scaled transverse kinetic energy, $(m_{\rm T}-m_0)/n_q$, where $n_q$ is the number of constituent quarks in the hadron, $m_0$ its mass, and $m_{\rm T}=\sqrt{p_{\rm T}^2+m_0^2}$. We find that the $D^0$ $v_2$ falls into the same universal trend as all other light hadrons~\cite{PHENIXmTscaling}, in particular for $(m_{\rm T}-m_0)/n_q$\,$<$\,1\,GeV/$c^2$. This suggests that charm quarks have gained significant flow through interactions with the sQGP medium in 10--40\% Au+Au collisions at \sNN = 200\,GeV.

The heavy quark-medium interaction is often characterized by a spatial diffusion coefficient $D_s$, or a dimensionless coefficient $2{\pi}TD_s$, where $T$ is the medium temperature~\cite{Moore:2004tg}. In Fig.~\ref{fig:v2CompareWithModel}, the measured $D^0$ $v_2$ in 0--80\% centrality collisions is compared with several model calculations
~\cite{PHSD,*PHSD2,SUBATECH,*SUBATECH2,*SUBATECHPrivateCom,Torino,*Torino2,hydroPang,*hydroPang2,*hydroPrivateCom,TAMU,*TAMU2,*TAMUPrivateCom,Duke,*DukePrivateCom,LBT}. Duke, LBT, PHSD, SUBATECH models and TAMU model with charm quark diffusion are able to describe our previously published $D^0$ $R_{\rm AA}$ result~\cite{StarRaa,PHSD,*PHSD2,LBT}.
Compared to the $v_2$ measurement, TAMU model with no charm quark diffusion does not reproduce the data, while the same model with charm quark diffusion turned on describes the data better~\cite{TAMU,*TAMU2,*TAMUPrivateCom}.
A 3D viscous event-by-event hydrodynamic simulation with $\eta/s$\,$=$\,0.12 using the AMPT initial condition and tuned to describe $v_2$ for light hadrons, predicts $D^0$ $v_2$ that is consistent with our data for $p_{\rm T}$\,$<$\,4\,\GeVc~\cite{hydroPang,*hydroPrivateCom}. This suggests that charm quarks have achieved thermal equilibrium in these collisions. 
We performed a statistical significance test for the consistency between our data and each model quantified by 
$\chi^2/\rm NDF$ and the $p$-value listed in Table~\ref{tab:chi2}. One can observe that the Duke model and TAMU model with no charm quark diffusion are inconsistent with our $v_2$ data, while other models describe the $v_2$ data in the measured $p_{\rm T}$ region. These models that can describe both the $R_{\rm AA}$ and $v_2$ data include the temperature--dependent charm diffusion coefficient $2{\pi}TD_s$ in the range of $\sim$2--12. 
$2{\pi}TD_s$ predicted by lattice QCD calculations fall in the same range~\cite{latticeBanerjee,latticeDing}. 
In addition to the different treatments of the charm--medium interactions, there are also various differences among these models, e.g. the initial state, the space-time description of the QGP evolution, the hadronization, and the interactions in the hadronic matter. More coherent model treatments of these aspects are needed in order to better interpret the information about charm-medium interaction, and provide a better constraint on 2$\pi TD_s$ using our $D^0$ $v_2$ measurement.

\begin{table}[htbp]
\caption{\label{tab:chi2}%
$D^0$ $v_2$ in 0--80\% centrality Au+Au collisions compared with model calculations,
 quantified by $\chi^2/\rm NDF$ and the $p$-value. 2$\pi TD_s$ values quoted are in the range of $T_c$ to 2$T_c$. $\chi^2/\rm NDF$ is calculated in the $p_{\rm T}$ range wherever the model calculation is available.
}
\begin{ruledtabular}
\begin{tabular}{lllr}
\textrm{compare with}&
\textrm{$2{\pi}TD_s$}&
\textrm{$\chi^{2}/\rm NDF$}&
\textrm{$p$-value}\\
\colrule
SUBATECH~\cite{SUBATECH,*SUBATECH2,*SUBATECHPrivateCom} & 2$-$4 & 15.2 / 8 & 0.06\\
TAMU c quark diff.~\cite{TAMU,*TAMU2,*TAMUPrivateCom} & 5$-$12 & 10.0 / 8 & 0.26\\
TAMU no c quark diff.~\cite{TAMU,*TAMU2,*TAMUPrivateCom} & - & 29.5 / 8 & 2\,$\times$\,$10^{-4}$\\
Duke~\cite{Duke} & 7 & 35.7 / 8 & 2\,$\times$\,$10^{-5}$ \\
LBT~\cite{LBT} & 3$-$6 & 11.1 / 8 & 0.19\\
PHSD~\cite{PHSD,*PHSD2} & 5$-$12 &  8.7 / 7 & 0.28\\
3D viscous hydro~\cite{hydroPang,*hydroPrivateCom} & - & 3.6 /  6 & 0.73\\
\end{tabular}
\end{ruledtabular}
\end{table}

In summary, the $D^0$ $v_2$ in Au+Au collisions at \sNN = 200\,GeV has been measured with the STAR detector using the Heavy Flavor Tracker, a newly installed high--resolution silicon detector. The measured $D^0$ $v_2$ follows the mass ordering at low $p_{\rm T}$ observed earlier. The $v_2/n_q$ of $D^0$ is consistent with that of other hadrons at $(m_{\rm T}-m_0)/n_q$\,$<$\,1\,GeV/$c^2$ in 10--40\% centrality collisions. A 3D viscous hydrodynamic model describes the $D^0$ $v_2$ for $p_{\rm T}$\,$<$\,4\,\GeVc. Our results suggest that charm quarks exhibit the same strong collective behavior as the light hadrons and may be close to thermal equilibrium in Au+Au collisions at \sNN = 200\,GeV.
Several theoretical calculations with temperature--dependent, dimensionless charm quark spatial diffusion coefficients ($2{\pi}TD_s$) in the range of $\sim$2--12 can simultaneously reproduce our $D^0$ $v_2$ result as well as the previously published STAR measurement of the $D^0$ nuclear modification factor. The charm quark diffusion coefficients from lattice QCD calculations are consistent with the same range~\cite{latticeBanerjee,latticeDing}.


\begin{acknowledgments}
We thank the RHIC Operations Group and RCF at BNL, the NERSC Center at LBNL, and the Open Science Grid consortium for providing resources and support. This work was supported in part by the Office of Nuclear Physics within the U.S. DOE Office of Science, the U.S. National Science Foundation, the Ministry of Education and Science of the Russian Federation, National Natural Science Foundation of China, Chinese Academy of Science, the Ministry of Science and Technology of China and the Chinese Ministry of Education, the National Research Foundation of Korea, GA and MSMT of the Czech Republic, Department of Atomic Energy and Department of Science and Technology of the Government of India; the National Science Centre of Poland, National Research Foundation, the Ministry of Science, Education and Sports of the Republic of Croatia, RosAtom of Russia and German Bundesministerium fur Bildung, Wissenschaft, Forschung and Technologie (BMBF) and the Helmholtz Association.
\end{acknowledgments}

\bibliography{draft_D0V2}

\begin{thebibliography}{59}%
\makeatletter
\providecommand \@ifxundefined [1]{%
 \@ifx{#1\undefined}
}%
\providecommand \@ifnum [1]{%
 \ifnum #1\expandafter \@firstoftwo
 \else \expandafter \@secondoftwo
 \fi
}%
\providecommand \@ifx [1]{%
 \ifx #1\expandafter \@firstoftwo
 \else \expandafter \@secondoftwo
 \fi
}%
\providecommand \natexlab [1]{#1}%
\providecommand \enquote  [1]{``#1''}%
\providecommand \bibnamefont  [1]{#1}%
\providecommand \bibfnamefont [1]{#1}%
\providecommand \citenamefont [1]{#1}%
\providecommand \href@noop [0]{\@secondoftwo}%
\providecommand \href [0]{\begingroup \@sanitize@url \@href}%
\providecommand \@href[1]{\@@startlink{#1}\@@href}%
\providecommand \@@href[1]{\endgroup#1\@@endlink}%
\providecommand \@sanitize@url [0]{\catcode `\\12\catcode `\$12\catcode
  `\&12\catcode `\#12\catcode `\^12\catcode `\_12\catcode `\%12\relax}%
\providecommand \@@startlink[1]{}%
\providecommand \@@endlink[0]{}%
\providecommand \url  [0]{\begingroup\@sanitize@url \@url }%
\providecommand \@url [1]{\endgroup\@href {#1}{\urlprefix }}%
\providecommand \urlprefix  [0]{URL }%
\providecommand \Eprint [0]{\href }%
\providecommand \doibase [0]{http://dx.doi.org/}%
\providecommand \selectlanguage [0]{\@gobble}%
\providecommand \bibinfo  [0]{\@secondoftwo}%
\providecommand \bibfield  [0]{\@secondoftwo}%
\providecommand \translation [1]{[#1]}%
\providecommand \BibitemOpen [0]{}%
\providecommand \bibitemStop [0]{}%
\providecommand \bibitemNoStop [0]{.\EOS\space}%
\providecommand \EOS [0]{\spacefactor3000\relax}%
\providecommand \BibitemShut  [1]{\csname bibitem#1\endcsname}%
\let\auto@bib@innerbib\@empty
\bibitem [{\citenamefont {Adams}\ \emph {et~al.}(2005)\citenamefont {Adams}
  \emph {et~al.}}]{StarWhitePaper}%
  \BibitemOpen
  \bibfield  {author} {\bibinfo {author} {\bibfnamefont {J.}~\bibnamefont
  {Adams}} \emph {et~al.} (\bibinfo {collaboration} {STAR}),\ }\href {\doibase
  10.1016/j.nuclphysa.2005.03.085} {\bibfield  {journal} {\bibinfo  {journal}
  {Nucl. Phys.}\ }\textbf {\bibinfo {volume} {A757}},\ \bibinfo {pages} {102}
  (\bibinfo {year} {2005})}\BibitemShut {NoStop}%
\bibitem [{\citenamefont {Adcox}\ \emph {et~al.}(2005)\citenamefont {Adcox}
  \emph {et~al.}}]{PhenixWhitePaper}%
  \BibitemOpen
  \bibfield  {author} {\bibinfo {author} {\bibfnamefont {K.}~\bibnamefont
  {Adcox}} \emph {et~al.} (\bibinfo {collaboration} {PHENIX}),\ }\href
  {\doibase 10.1016/j.nuclphysa.2005.03.086} {\bibfield  {journal} {\bibinfo
  {journal} {Nucl. Phys.}\ }\textbf {\bibinfo {volume} {A757}},\ \bibinfo
  {pages} {184} (\bibinfo {year} {2005})}\BibitemShut {NoStop}%
\bibitem [{\citenamefont {Muller}\ \emph {et~al.}(2012)\citenamefont {Muller},
  \citenamefont {Schukraft},\ and\ \citenamefont {Wyslouch}}]{LhcSummary}%
  \BibitemOpen
  \bibfield  {author} {\bibinfo {author} {\bibfnamefont {B.}~\bibnamefont
  {Muller}}, \bibinfo {author} {\bibfnamefont {J.}~\bibnamefont {Schukraft}}, \
  and\ \bibinfo {author} {\bibfnamefont {B.}~\bibnamefont {Wyslouch}},\ }\href
  {\doibase 10.1146/annurev-nucl-102711-094910} {\bibfield  {journal} {\bibinfo
   {journal} {Ann. Rev. Nucl. Part. Sci.}\ }\textbf {\bibinfo {volume} {62}},\
  \bibinfo {pages} {361} (\bibinfo {year} {2012})}\BibitemShut {NoStop}%
\bibitem [{\citenamefont {Gale}\ \emph {et~al.}(2013)\citenamefont {Gale},
  \citenamefont {Jeon}, \citenamefont {Schenke}, \citenamefont {Tribedy},\ and\
  \citenamefont {Venugopalan}}]{Gale:2012rq}%
  \BibitemOpen
  \bibfield  {author} {\bibinfo {author} {\bibfnamefont {C.}~\bibnamefont
  {Gale}}, \bibinfo {author} {\bibfnamefont {S.}~\bibnamefont {Jeon}}, \bibinfo
  {author} {\bibfnamefont {B.}~\bibnamefont {Schenke}}, \bibinfo {author}
  {\bibfnamefont {P.}~\bibnamefont {Tribedy}}, \ and\ \bibinfo {author}
  {\bibfnamefont {R.}~\bibnamefont {Venugopalan}},\ }\href {\doibase
  10.1103/PhysRevLett.110.012302} {\bibfield  {journal} {\bibinfo  {journal}
  {Phys. Rev. Lett.}\ }\textbf {\bibinfo {volume} {110}},\ \bibinfo {pages}
  {012302} (\bibinfo {year} {2013})}\BibitemShut {NoStop}%
\bibitem [{\citenamefont {Moore}\ and\ \citenamefont
  {Teaney}(2005)}]{Moore:2004tg}%
  \BibitemOpen
  \bibfield  {author} {\bibinfo {author} {\bibfnamefont {G.~D.}\ \bibnamefont
  {Moore}}\ and\ \bibinfo {author} {\bibfnamefont {D.}~\bibnamefont {Teaney}},\
  }\href {\doibase 10.1103/PhysRevC.71.064904} {\bibfield  {journal} {\bibinfo
  {journal} {Phys. Rev.}\ }\textbf {\bibinfo {volume} {C71}},\ \bibinfo {pages}
  {064904} (\bibinfo {year} {2005})}\BibitemShut {NoStop}%
\bibitem [{\citenamefont {Rapp}\ and\ \citenamefont {van
  Hees}(2008)}]{Rapp:2008qc}%
  \BibitemOpen
  \bibfield  {author} {\bibinfo {author} {\bibfnamefont {R.}~\bibnamefont
  {Rapp}}\ and\ \bibinfo {author} {\bibfnamefont {H.}~\bibnamefont {van
  Hees}},\ }\href@noop {} {\  (\bibinfo {year} {2008})},\ \Eprint
  {http://arxiv.org/abs/0803.0901} {arXiv:0803.0901 [hep-ph]} \BibitemShut
  {NoStop}%
\bibitem [{\citenamefont {Wang}\ and\ \citenamefont
  {Gyulassy}(1992)}]{Wang:1991xy}%
  \BibitemOpen
  \bibfield  {author} {\bibinfo {author} {\bibfnamefont {X.-N.}\ \bibnamefont
  {Wang}}\ and\ \bibinfo {author} {\bibfnamefont {M.}~\bibnamefont
  {Gyulassy}},\ }\href {\doibase 10.1103/PhysRevLett.68.1480} {\bibfield
  {journal} {\bibinfo  {journal} {Phys. Rev. Lett.}\ }\textbf {\bibinfo
  {volume} {68}},\ \bibinfo {pages} {1480} (\bibinfo {year}
  {1992})}\BibitemShut {NoStop}%
\bibitem [{\citenamefont {Poskanzer}\ and\ \citenamefont
  {Voloshin}(1998)}]{Poskanzer98}%
  \BibitemOpen
  \bibfield  {author} {\bibinfo {author} {\bibfnamefont {A.~M.}\ \bibnamefont
  {Poskanzer}}\ and\ \bibinfo {author} {\bibfnamefont {S.~A.}\ \bibnamefont
  {Voloshin}},\ }\href {\doibase 10.1103/PhysRevC.58.1671} {\bibfield
  {journal} {\bibinfo  {journal} {Phys. Rev.}\ }\textbf {\bibinfo {volume}
  {C58}},\ \bibinfo {pages} {1671} (\bibinfo {year} {1998})}\BibitemShut
  {NoStop}%
\bibitem [{\citenamefont {Andronic}\ \emph {et~al.}(2016)\citenamefont
  {Andronic} \emph {et~al.}}]{HFSummary}%
  \BibitemOpen
  \bibfield  {author} {\bibinfo {author} {\bibfnamefont {A.}~\bibnamefont
  {Andronic}} \emph {et~al.},\ }\href {\doibase 10.1140/epjc/s10052-015-3819-5}
  {\bibfield  {journal} {\bibinfo  {journal} {Eur. Phys. J.}\ }\textbf
  {\bibinfo {volume} {C76}},\ \bibinfo {pages} {107} (\bibinfo {year}
  {2016})}\BibitemShut {NoStop}%
\bibitem [{\citenamefont {Adamczyk}\ \emph
  {et~al.}(2014{\natexlab{a}})\citenamefont {Adamczyk} \emph
  {et~al.}}]{StarRaa}%
  \BibitemOpen
  \bibfield  {author} {\bibinfo {author} {\bibfnamefont {L.}~\bibnamefont
  {Adamczyk}} \emph {et~al.} (\bibinfo {collaboration} {STAR}),\ }\href
  {\doibase 10.1103/PhysRevLett.113.142301} {\bibfield  {journal} {\bibinfo
  {journal} {Phys. Rev. Lett.}\ }\textbf {\bibinfo {volume} {113}},\ \bibinfo
  {pages} {142301} (\bibinfo {year} {2014}{\natexlab{a}})}\BibitemShut
  {NoStop}%
\bibitem [{\citenamefont {Abelev}\ \emph {et~al.}(2012)\citenamefont {Abelev}
  \emph {et~al.}}]{AliceRaa}%
  \BibitemOpen
  \bibfield  {author} {\bibinfo {author} {\bibfnamefont {B.}~\bibnamefont
  {Abelev}} \emph {et~al.} (\bibinfo {collaboration} {ALICE}),\ }\href
  {\doibase 10.1007/JHEP09(2012)112} {\bibfield  {journal} {\bibinfo  {journal}
  {JHEP}\ }\textbf {\bibinfo {volume} {09}},\ \bibinfo {pages} {112} (\bibinfo
  {year} {2012})}\BibitemShut {NoStop}%
\bibitem [{\citenamefont {Adam}\ \emph {et~al.}(2016)\citenamefont {Adam} \emph
  {et~al.}}]{AliceRaa2}%
  \BibitemOpen
  \bibfield  {author} {\bibinfo {author} {\bibfnamefont {J.}~\bibnamefont
  {Adam}} \emph {et~al.} (\bibinfo {collaboration} {ALICE}),\ }\href {\doibase
  10.1007/JHEP03(2016)081} {\bibfield  {journal} {\bibinfo  {journal} {JHEP}\
  }\textbf {\bibinfo {volume} {03}},\ \bibinfo {pages} {081} (\bibinfo {year}
  {2016})}\BibitemShut {NoStop}%
\bibitem [{\citenamefont {Abelev}\ \emph {et~al.}(2013)\citenamefont {Abelev}
  \emph {et~al.}}]{AliceV2}%
  \BibitemOpen
  \bibfield  {author} {\bibinfo {author} {\bibfnamefont {B.}~\bibnamefont
  {Abelev}} \emph {et~al.} (\bibinfo {collaboration} {ALICE}),\ }\href
  {\doibase 10.1103/PhysRevLett.111.102301} {\bibfield  {journal} {\bibinfo
  {journal} {Phys. Rev. Lett.}\ }\textbf {\bibinfo {volume} {111}},\ \bibinfo
  {pages} {102301} (\bibinfo {year} {2013})}\BibitemShut {NoStop}%
\bibitem [{\citenamefont {Abelev}\ \emph {et~al.}(2014)\citenamefont {Abelev}
  \emph {et~al.}}]{AliceV22}%
  \BibitemOpen
  \bibfield  {author} {\bibinfo {author} {\bibfnamefont {B.~B.}\ \bibnamefont
  {Abelev}} \emph {et~al.} (\bibinfo {collaboration} {ALICE}),\ }\href
  {\doibase 10.1103/PhysRevC.90.034904} {\bibfield  {journal} {\bibinfo
  {journal} {Phys. Rev.}\ }\textbf {\bibinfo {volume} {C90}},\ \bibinfo {pages}
  {034904} (\bibinfo {year} {2014})}\BibitemShut {NoStop}%
\bibitem [{\citenamefont {Adare}\ \emph {et~al.}(2007)\citenamefont {Adare}
  \emph {et~al.}}]{PhenixNpe}%
  \BibitemOpen
  \bibfield  {author} {\bibinfo {author} {\bibfnamefont {A.}~\bibnamefont
  {Adare}} \emph {et~al.} (\bibinfo {collaboration} {PHENIX}),\ }\href
  {\doibase 10.1103/PhysRevLett.98.172301} {\bibfield  {journal} {\bibinfo
  {journal} {Phys. Rev. Lett.}\ }\textbf {\bibinfo {volume} {98}},\ \bibinfo
  {pages} {172301} (\bibinfo {year} {2007})}\BibitemShut {NoStop}%
\bibitem [{\citenamefont {Adare}\ \emph {et~al.}(2011)\citenamefont {Adare}
  \emph {et~al.}}]{PhenixNpe2}%
  \BibitemOpen
  \bibfield  {author} {\bibinfo {author} {\bibfnamefont {A.}~\bibnamefont
  {Adare}} \emph {et~al.} (\bibinfo {collaboration} {PHENIX}),\ }\href
  {\doibase 10.1103/PhysRevC.84.044905} {\bibfield  {journal} {\bibinfo
  {journal} {Phys. Rev.}\ }\textbf {\bibinfo {volume} {C84}},\ \bibinfo {pages}
  {044905} (\bibinfo {year} {2011})}\BibitemShut {NoStop}%
\bibitem [{\citenamefont {Adamczyk}\ \emph
  {et~al.}(2014{\natexlab{b}})\citenamefont {Adamczyk} \emph
  {et~al.}}]{STARNpe}%
  \BibitemOpen
  \bibfield  {author} {\bibinfo {author} {\bibfnamefont {L.}~\bibnamefont
  {Adamczyk}} \emph {et~al.} (\bibinfo {collaboration} {STAR}),\ }\href@noop {}
  {\  (\bibinfo {year} {2014}{\natexlab{b}})},\ \Eprint
  {http://arxiv.org/abs/1405.6348} {arXiv:1405.6348 [hep-ex]} \BibitemShut
  {NoStop}%
\bibitem [{\citenamefont {Berrehrah}\ \emph {et~al.}(2014)\citenamefont
  {Berrehrah}, \citenamefont {Gossiaux}, \citenamefont {Aichelin},
  \citenamefont {Cassing}, \citenamefont {Torres-Rincon},\ and\ \citenamefont
  {Bratkovskaya}}]{PHSD}%
  \BibitemOpen
  \bibfield  {author} {\bibinfo {author} {\bibfnamefont {H.}~\bibnamefont
  {Berrehrah}}, \bibinfo {author} {\bibfnamefont {P.~B.}\ \bibnamefont
  {Gossiaux}}, \bibinfo {author} {\bibfnamefont {J.}~\bibnamefont {Aichelin}},
  \bibinfo {author} {\bibfnamefont {W.}~\bibnamefont {Cassing}}, \bibinfo
  {author} {\bibfnamefont {J.~M.}\ \bibnamefont {Torres-Rincon}}, \ and\
  \bibinfo {author} {\bibfnamefont {E.}~\bibnamefont {Bratkovskaya}},\ }\href
  {\doibase 10.1103/PhysRevC.90.051901} {\bibfield  {journal} {\bibinfo
  {journal} {Phys. Rev.}\ }\textbf {\bibinfo {volume} {C90}},\ \bibinfo {pages}
  {051901} (\bibinfo {year} {2014})}\BibitemShut {NoStop}%
\bibitem [{\citenamefont {Song}\ \emph {et~al.}(2015)\citenamefont {Song},
  \citenamefont {Berrehrah}, \citenamefont {Cebrera}, \citenamefont
  {Torres-Rincon}, \citenamefont {Tolos}, \citenamefont {Cassing},\ and\
  \citenamefont {Bratkovskaya}}]{PHSD2}%
  \BibitemOpen
  \bibfield  {author} {\bibinfo {author} {\bibfnamefont {T.}~\bibnamefont
  {Song}}, \bibinfo {author} {\bibfnamefont {H.}~\bibnamefont {Berrehrah}},
  \bibinfo {author} {\bibfnamefont {D.}~\bibnamefont {Cebrera}}, \bibinfo
  {author} {\bibfnamefont {J.}~\bibnamefont {Torres-Rincon}}, \bibinfo {author}
  {\bibfnamefont {L.}~\bibnamefont {Tolos}}, \bibinfo {author} {\bibfnamefont
  {W.}~\bibnamefont {Cassing}}, \ and\ \bibinfo {author} {\bibfnamefont
  {E.}~\bibnamefont {Bratkovskaya}},\ }\href {\doibase
  10.1103/PhysRevC.92.014910} {\bibfield  {journal} {\bibinfo  {journal} {Phys.
  Rev.}\ }\textbf {\bibinfo {volume} {C92}},\ \bibinfo {pages} {014910}
  (\bibinfo {year} {2015})}\BibitemShut {NoStop}%
\bibitem [{\citenamefont {Ozvenchuk}\ \emph {et~al.}(2014)\citenamefont
  {Ozvenchuk}, \citenamefont {Torres-Rincon}, \citenamefont {Gossiaux},
  \citenamefont {Aichelin},\ and\ \citenamefont {Tolos}}]{SUBATECH}%
  \BibitemOpen
  \bibfield  {author} {\bibinfo {author} {\bibfnamefont {V.}~\bibnamefont
  {Ozvenchuk}}, \bibinfo {author} {\bibfnamefont {J.}~\bibnamefont
  {Torres-Rincon}}, \bibinfo {author} {\bibfnamefont {P.}~\bibnamefont
  {Gossiaux}}, \bibinfo {author} {\bibfnamefont {J.}~\bibnamefont {Aichelin}},
  \ and\ \bibinfo {author} {\bibfnamefont {L.}~\bibnamefont {Tolos}},\ }\href
  {\doibase 10.1103/PhysRevC.90.054909} {\bibfield  {journal} {\bibinfo
  {journal} {Phys. Rev.}\ }\textbf {\bibinfo {volume} {C90}},\ \bibinfo {pages}
  {054909} (\bibinfo {year} {2014})}\BibitemShut {NoStop}%
\bibitem [{\citenamefont {Nahrgang}\ \emph {et~al.}(2015)\citenamefont
  {Nahrgang}, \citenamefont {Aichelin}, \citenamefont {Bass}, \citenamefont
  {Gossiaux},\ and\ \citenamefont {Werner}}]{SUBATECH2}%
  \BibitemOpen
  \bibfield  {author} {\bibinfo {author} {\bibfnamefont {M.}~\bibnamefont
  {Nahrgang}}, \bibinfo {author} {\bibfnamefont {J.}~\bibnamefont {Aichelin}},
  \bibinfo {author} {\bibfnamefont {S.}~\bibnamefont {Bass}}, \bibinfo {author}
  {\bibfnamefont {P.~B.}\ \bibnamefont {Gossiaux}}, \ and\ \bibinfo {author}
  {\bibfnamefont {K.}~\bibnamefont {Werner}},\ }\href {\doibase
  10.1103/PhysRevC.91.014904} {\bibfield  {journal} {\bibinfo  {journal} {Phys.
  Rev.}\ }\textbf {\bibinfo {volume} {C91}},\ \bibinfo {pages} {014904}
  (\bibinfo {year} {2015})}\BibitemShut {NoStop}%
\bibitem [{SUB()}]{SUBATECHPrivateCom}%
  \BibitemOpen
  \href@noop {} {}\bibinfo {howpublished} {and private
  communication}\BibitemShut {NoStop}%
\bibitem [{\citenamefont {Alberico}\ \emph {et~al.}(2011)\citenamefont
  {Alberico}, \citenamefont {Beraudo}, \citenamefont {De~Pace}, \citenamefont
  {Molinari}, \citenamefont {Monteno}, \citenamefont {Nardi},\ and\
  \citenamefont {Prino}}]{Torino}%
  \BibitemOpen
  \bibfield  {author} {\bibinfo {author} {\bibfnamefont {W.~M.}\ \bibnamefont
  {Alberico}}, \bibinfo {author} {\bibfnamefont {A.}~\bibnamefont {Beraudo}},
  \bibinfo {author} {\bibfnamefont {A.}~\bibnamefont {De~Pace}}, \bibinfo
  {author} {\bibfnamefont {A.}~\bibnamefont {Molinari}}, \bibinfo {author}
  {\bibfnamefont {M.}~\bibnamefont {Monteno}}, \bibinfo {author} {\bibfnamefont
  {M.}~\bibnamefont {Nardi}}, \ and\ \bibinfo {author} {\bibfnamefont
  {F.}~\bibnamefont {Prino}},\ }\href {\doibase 10.1140/epjc/s10052-011-1666-6}
  {\bibfield  {journal} {\bibinfo  {journal} {Eur. Phys. J.}\ }\textbf
  {\bibinfo {volume} {C71}},\ \bibinfo {pages} {1666} (\bibinfo {year}
  {2011})}\BibitemShut {NoStop}%
\bibitem [{\citenamefont {Beraudo}\ \emph {et~al.}(2015)\citenamefont
  {Beraudo}, \citenamefont {De~Pace}, \citenamefont {Monteno}, \citenamefont
  {Nardi},\ and\ \citenamefont {Prino}}]{Torino2}%
  \BibitemOpen
  \bibfield  {author} {\bibinfo {author} {\bibfnamefont {A.}~\bibnamefont
  {Beraudo}}, \bibinfo {author} {\bibfnamefont {A.}~\bibnamefont {De~Pace}},
  \bibinfo {author} {\bibfnamefont {M.}~\bibnamefont {Monteno}}, \bibinfo
  {author} {\bibfnamefont {M.}~\bibnamefont {Nardi}}, \ and\ \bibinfo {author}
  {\bibfnamefont {F.}~\bibnamefont {Prino}},\ }\href {\doibase
  10.1140/epjc/s10052-015-3336-6} {\bibfield  {journal} {\bibinfo  {journal}
  {Eur. Phys. J.}\ }\textbf {\bibinfo {volume} {C75}},\ \bibinfo {pages} {121}
  (\bibinfo {year} {2015})}\BibitemShut {NoStop}%
\bibitem [{\citenamefont {Cao}\ \emph {et~al.}(2015)\citenamefont {Cao},
  \citenamefont {Qin},\ and\ \citenamefont {Bass}}]{Duke}%
  \BibitemOpen
  \bibfield  {author} {\bibinfo {author} {\bibfnamefont {S.~S.}\ \bibnamefont
  {Cao}}, \bibinfo {author} {\bibfnamefont {G.}~\bibnamefont {Qin}}, \ and\
  \bibinfo {author} {\bibfnamefont {S.~A.}\ \bibnamefont {Bass}},\ }\href
  {\doibase 10.1103/PhysRevC.92.024907} {\bibfield  {journal} {\bibinfo
  {journal} {Phys. Rev.}\ }\textbf {\bibinfo {volume} {C92}},\ \bibinfo {pages}
  {024907} (\bibinfo {year} {2015})}\BibitemShut {NoStop}%
\bibitem [{Duk()}]{DukePrivateCom}%
  \BibitemOpen
  \href@noop {} {}\bibinfo {howpublished} {and private
  communication}\BibitemShut {NoStop}%
\bibitem [{\citenamefont {He}\ \emph {et~al.}(2012)\citenamefont {He},
  \citenamefont {Fries},\ and\ \citenamefont {Rapp}}]{TAMU}%
  \BibitemOpen
  \bibfield  {author} {\bibinfo {author} {\bibfnamefont {M.}~\bibnamefont
  {He}}, \bibinfo {author} {\bibfnamefont {R.~J.}\ \bibnamefont {Fries}}, \
  and\ \bibinfo {author} {\bibfnamefont {R.}~\bibnamefont {Rapp}},\ }\href
  {\doibase 10.1103/PhysRevC.86.014903} {\bibfield  {journal} {\bibinfo
  {journal} {Phys. Rev.}\ }\textbf {\bibinfo {volume} {C86}},\ \bibinfo {pages}
  {014903} (\bibinfo {year} {2012})}\BibitemShut {NoStop}%
\bibitem [{\citenamefont {He}\ \emph {et~al.}(2013)\citenamefont {He},
  \citenamefont {Fries},\ and\ \citenamefont {Rapp}}]{TAMU2}%
  \BibitemOpen
  \bibfield  {author} {\bibinfo {author} {\bibfnamefont {M.}~\bibnamefont
  {He}}, \bibinfo {author} {\bibfnamefont {R.~J.}\ \bibnamefont {Fries}}, \
  and\ \bibinfo {author} {\bibfnamefont {R.}~\bibnamefont {Rapp}},\ }\href
  {\doibase 10.1103/PhysRevLett.110.112301} {\bibfield  {journal} {\bibinfo
  {journal} {Phys. Rev. Lett.}\ }\textbf {\bibinfo {volume} {110}},\ \bibinfo
  {pages} {112301} (\bibinfo {year} {2013})}\BibitemShut {NoStop}%
\bibitem [{TAM()}]{TAMUPrivateCom}%
  \BibitemOpen
  \href@noop {} {}\bibinfo {howpublished} {and private
  communication}\BibitemShut {NoStop}%
\bibitem [{\citenamefont {Cao}\ \emph {et~al.}(2016)\citenamefont {Cao},
  \citenamefont {Luo}, \citenamefont {Qin},\ and\ \citenamefont {Wang}}]{LBT}%
  \BibitemOpen
  \bibfield  {author} {\bibinfo {author} {\bibfnamefont {S.}~\bibnamefont
  {Cao}}, \bibinfo {author} {\bibfnamefont {T.}~\bibnamefont {Luo}}, \bibinfo
  {author} {\bibfnamefont {G.-Y.}\ \bibnamefont {Qin}}, \ and\ \bibinfo
  {author} {\bibfnamefont {X.-N.}\ \bibnamefont {Wang}},\ }\href {\doibase
  10.1103/PhysRevC.94.014909} {\bibfield  {journal} {\bibinfo  {journal} {Phys.
  Rev.}\ }\textbf {\bibinfo {volume} {C94}},\ \bibinfo {pages} {014909}
  (\bibinfo {year} {2016})}\BibitemShut {NoStop}%
\bibitem [{\citenamefont {Uphoff}\ \emph {et~al.}(2012)\citenamefont {Uphoff},
  \citenamefont {Fochler}, \citenamefont {Xu},\ and\ \citenamefont
  {Greiner}}]{BAMPS}%
  \BibitemOpen
  \bibfield  {author} {\bibinfo {author} {\bibfnamefont {J.}~\bibnamefont
  {Uphoff}}, \bibinfo {author} {\bibfnamefont {O.}~\bibnamefont {Fochler}},
  \bibinfo {author} {\bibfnamefont {Z.}~\bibnamefont {Xu}}, \ and\ \bibinfo
  {author} {\bibfnamefont {C.}~\bibnamefont {Greiner}},\ }\href {\doibase
  10.1016/j.physletb.2012.09.069} {\bibfield  {journal} {\bibinfo  {journal}
  {Phys. Lett.}\ }\textbf {\bibinfo {volume} {B717}},\ \bibinfo {pages} {430}
  (\bibinfo {year} {2012})}\BibitemShut {NoStop}%
\bibitem [{\citenamefont {Uphoff}\ \emph {et~al.}(2014)\citenamefont {Uphoff},
  \citenamefont {Fochler}, \citenamefont {Xu},\ and\ \citenamefont
  {Greiner}}]{BAMPS2}%
  \BibitemOpen
  \bibfield  {author} {\bibinfo {author} {\bibfnamefont {J.}~\bibnamefont
  {Uphoff}}, \bibinfo {author} {\bibfnamefont {O.}~\bibnamefont {Fochler}},
  \bibinfo {author} {\bibfnamefont {Z.}~\bibnamefont {Xu}}, \ and\ \bibinfo
  {author} {\bibfnamefont {C.}~\bibnamefont {Greiner}},\ }\href {\doibase
  10.1016/j.nuclphysa.2014.07.024} {\bibfield  {journal} {\bibinfo  {journal}
  {Nucl. Phys.}\ }\textbf {\bibinfo {volume} {A932}},\ \bibinfo {pages} {247}
  (\bibinfo {year} {2014})}\BibitemShut {NoStop}%
\bibitem [{\citenamefont {Sharma}\ \emph {et~al.}(2009)\citenamefont {Sharma},
  \citenamefont {Vitev},\ and\ \citenamefont {Zhang}}]{LANL}%
  \BibitemOpen
  \bibfield  {author} {\bibinfo {author} {\bibfnamefont {R.}~\bibnamefont
  {Sharma}}, \bibinfo {author} {\bibfnamefont {I.}~\bibnamefont {Vitev}}, \
  and\ \bibinfo {author} {\bibfnamefont {B.-W.}\ \bibnamefont {Zhang}},\ }\href
  {\doibase 10.1103/PhysRevC.80.054902} {\bibfield  {journal} {\bibinfo
  {journal} {Phys. Rev.}\ }\textbf {\bibinfo {volume} {C80}},\ \bibinfo {pages}
  {054902} (\bibinfo {year} {2009})}\BibitemShut {NoStop}%
\bibitem [{\citenamefont {Beavis}\ \emph {et~al.}(2011)\citenamefont {Beavis}
  \emph {et~al.}}]{HFT}%
  \BibitemOpen
  \bibfield  {author} {\bibinfo {author} {\bibfnamefont {D.}~\bibnamefont
  {Beavis}} \emph {et~al.},\ }\href
  {https://drupal.star.bnl.gov/STAR/starnotes/public/sn0600} {\enquote
  {\bibinfo {title} {The star heavy flavor tracker technical design report},}\
  }\bibinfo {howpublished}
  {{\burl{https://drupal.star.bnl.gov/STAR/starnotes/public/sn0600}}} (\bibinfo
  {year} {2011})\BibitemShut {NoStop}%
\bibitem [{\citenamefont {Qiu}(2014)}]{HFTQM14}%
  \BibitemOpen
  \bibfield  {author} {\bibinfo {author} {\bibfnamefont {H.}~\bibnamefont
  {Qiu}} (\bibinfo {collaboration} {STAR}),\ }\href {\doibase
  10.1016/j.nuclphysa.2014.08.056} {\bibfield  {journal} {\bibinfo  {journal}
  {Nucl. Phys.}\ }\textbf {\bibinfo {volume} {A931}},\ \bibinfo {pages} {1141}
  (\bibinfo {year} {2014})}\BibitemShut {NoStop}%
\bibitem [{\citenamefont {Llope}\ \emph {et~al.}(2004)\citenamefont {Llope}
  \emph {et~al.}}]{VPD}%
  \BibitemOpen
  \bibfield  {author} {\bibinfo {author} {\bibfnamefont {W.~J.}\ \bibnamefont
  {Llope}} \emph {et~al.},\ }\href {\doibase 10.1016/j.nima.2003.11.414}
  {\bibfield  {journal} {\bibinfo  {journal} {Nucl. Instrum. Meth.}\ }\textbf
  {\bibinfo {volume} {A522}},\ \bibinfo {pages} {252} (\bibinfo {year}
  {2004})}\BibitemShut {NoStop}%
\bibitem [{\citenamefont {Abelev}\ \emph {et~al.}(2009)\citenamefont {Abelev}
  \emph {et~al.}}]{MCGlauber}%
  \BibitemOpen
  \bibfield  {author} {\bibinfo {author} {\bibfnamefont {B.}~\bibnamefont
  {Abelev}} \emph {et~al.} (\bibinfo {collaboration} {STAR}),\ }\href {\doibase
  10.1103/PhysRevC.79.034909} {\bibfield  {journal} {\bibinfo  {journal} {Phys.
  Rev.}\ }\textbf {\bibinfo {volume} {C79}},\ \bibinfo {pages} {034909}
  (\bibinfo {year} {2009})}\BibitemShut {NoStop}%
\bibitem [{\citenamefont {Olive}\ \emph {et~al.}(2014)\citenamefont {Olive}
  \emph {et~al.}}]{PDG}%
  \BibitemOpen
  \bibfield  {author} {\bibinfo {author} {\bibfnamefont {K.}~\bibnamefont
  {Olive}} \emph {et~al.},\ }\href@noop {} {\bibfield  {journal} {\bibinfo
  {journal} {Chin. Phys.}\ }\textbf {\bibinfo {volume} {C38}},\ \bibinfo
  {pages} {090001} (\bibinfo {year} {2014})}\BibitemShut {NoStop}%
\bibitem [{\citenamefont {Anderson}\ \emph {et~al.}(2003)\citenamefont
  {Anderson} \emph {et~al.}}]{TPC}%
  \BibitemOpen
  \bibfield  {author} {\bibinfo {author} {\bibfnamefont {M.}~\bibnamefont
  {Anderson}} \emph {et~al.},\ }\href {\doibase 10.1016/S0168-9002(02)01964-2}
  {\bibfield  {journal} {\bibinfo  {journal} {Nucl. Instrum. Meth.}\ }\textbf
  {\bibinfo {volume} {A499}},\ \bibinfo {pages} {659} (\bibinfo {year}
  {2003})}\BibitemShut {NoStop}%
\bibitem [{\citenamefont {Llope}(2012)}]{TOF}%
  \BibitemOpen
  \bibfield  {author} {\bibinfo {author} {\bibfnamefont {W.~J.}\ \bibnamefont
  {Llope}} (\bibinfo {collaboration} {STAR}),\ }\href {\doibase
  10.1016/j.nima.2010.07.086} {\bibfield  {journal} {\bibinfo  {journal} {Nucl.
  Instrum. Meth.}\ }\textbf {\bibinfo {volume} {A661}},\ \bibinfo {pages}
  {S110} (\bibinfo {year} {2012})}\BibitemShut {NoStop}%
\bibitem [{\citenamefont {Hocker}\ \emph {et~al.}(2007)\citenamefont {Hocker}
  \emph {et~al.}}]{TMVA}%
  \BibitemOpen
  \bibfield  {author} {\bibinfo {author} {\bibfnamefont {A.}~\bibnamefont
  {Hocker}} \emph {et~al.},\ }\href@noop {} {\bibfield  {journal} {\bibinfo
  {journal} {PoS}\ }\textbf {\bibinfo {volume} {ACAT}},\ \bibinfo {pages} {040}
  (\bibinfo {year} {2007})}\BibitemShut {NoStop}%
\bibitem [{\citenamefont {Borghini}\ \emph {et~al.}(2001)\citenamefont
  {Borghini}, \citenamefont {Dinh},\ and\ \citenamefont
  {Ollitrault}}]{correlationMethod}%
  \BibitemOpen
  \bibfield  {author} {\bibinfo {author} {\bibfnamefont {N.}~\bibnamefont
  {Borghini}}, \bibinfo {author} {\bibfnamefont {P.~M.}\ \bibnamefont {Dinh}},
  \ and\ \bibinfo {author} {\bibfnamefont {J.-Y.}\ \bibnamefont {Ollitrault}},\
  }\href {\doibase 10.1103/PhysRevC.63.054906} {\bibfield  {journal} {\bibinfo
  {journal} {Phys. Rev.}\ }\textbf {\bibinfo {volume} {C63}},\ \bibinfo {pages}
  {054906} (\bibinfo {year} {2001})}\BibitemShut {NoStop}%
\bibitem [{\citenamefont {Bilandzic}\ \emph {et~al.}(2011)\citenamefont
  {Bilandzic}, \citenamefont {Snellings},\ and\ \citenamefont
  {Voloshin}}]{correlationMethod2}%
  \BibitemOpen
  \bibfield  {author} {\bibinfo {author} {\bibfnamefont {A.}~\bibnamefont
  {Bilandzic}}, \bibinfo {author} {\bibfnamefont {R.}~\bibnamefont
  {Snellings}}, \ and\ \bibinfo {author} {\bibfnamefont {S.}~\bibnamefont
  {Voloshin}},\ }\href {\doibase 10.1103/PhysRevC.83.044913} {\bibfield
  {journal} {\bibinfo  {journal} {Phys. Rev.}\ }\textbf {\bibinfo {volume}
  {C83}},\ \bibinfo {pages} {044913} (\bibinfo {year} {2011})}\BibitemShut
  {NoStop}%
\bibitem [{\citenamefont {Masui}\ \emph {et~al.}(2016)\citenamefont {Masui},
  \citenamefont {Schmah},\ and\ \citenamefont {Poskanzer}}]{Masui:2012zh}%
  \BibitemOpen
  \bibfield  {author} {\bibinfo {author} {\bibfnamefont {H.}~\bibnamefont
  {Masui}}, \bibinfo {author} {\bibfnamefont {A.}~\bibnamefont {Schmah}}, \
  and\ \bibinfo {author} {\bibfnamefont {A.~M.}\ \bibnamefont {Poskanzer}},\
  }\href {\doibase 10.1016/j.nima.2016.07.037} {\bibfield  {journal} {\bibinfo
  {journal} {Nucl. Instrum. Meth.}\ }\textbf {\bibinfo {volume} {A833}},\
  \bibinfo {pages} {181} (\bibinfo {year} {2016})}\BibitemShut {NoStop}%
\bibitem [{\citenamefont {Adamczyk}\ \emph {et~al.}(2012)\citenamefont
  {Adamczyk} \emph {et~al.}}]{Adamczyk:2012af}%
  \BibitemOpen
  \bibfield  {author} {\bibinfo {author} {\bibfnamefont {L.}~\bibnamefont
  {Adamczyk}} \emph {et~al.} (\bibinfo {collaboration} {STAR}),\ }\href
  {\doibase 10.1103/PhysRevD.86.072013} {\bibfield  {journal} {\bibinfo
  {journal} {Phys. Rev.}\ }\textbf {\bibinfo {volume} {D86}},\ \bibinfo {pages}
  {072013} (\bibinfo {year} {2012})},\ \Eprint {http://arxiv.org/abs/1204.4244}
  {arXiv:1204.4244 [nucl-ex]} \BibitemShut {NoStop}%
\bibitem [{\citenamefont {Abelev}\ \emph {et~al.}(2008)\citenamefont {Abelev}
  \emph {et~al.}}]{ksV2}%
  \BibitemOpen
  \bibfield  {author} {\bibinfo {author} {\bibfnamefont {B.~I.}\ \bibnamefont
  {Abelev}} \emph {et~al.} (\bibinfo {collaboration} {STAR}),\ }\href {\doibase
  10.1103/PhysRevC.77.054901} {\bibfield  {journal} {\bibinfo  {journal} {Phys.
  Rev.}\ }\textbf {\bibinfo {volume} {C77}},\ \bibinfo {pages} {054901}
  (\bibinfo {year} {2008})}\BibitemShut {NoStop}%
\bibitem [{\citenamefont {Adamczyk}\ \emph {et~al.}(2016)\citenamefont
  {Adamczyk} \emph {et~al.}}]{phiOmegaV2}%
  \BibitemOpen
  \bibfield  {author} {\bibinfo {author} {\bibfnamefont {L.}~\bibnamefont
  {Adamczyk}} \emph {et~al.} (\bibinfo {collaboration} {STAR}),\ }\href
  {\doibase 10.1103/PhysRevLett.116.062301} {\bibfield  {journal} {\bibinfo
  {journal} {Phys. Rev. Lett.}\ }\textbf {\bibinfo {volume} {116}},\ \bibinfo
  {pages} {062301} (\bibinfo {year} {2016})}\BibitemShut {NoStop}%
\bibitem [{\citenamefont {Adams}\ \emph {et~al.}(2004)\citenamefont {Adams}
  \emph {et~al.}}]{Adams:2004wz}%
  \BibitemOpen
  \bibfield  {author} {\bibinfo {author} {\bibfnamefont {J.}~\bibnamefont
  {Adams}} \emph {et~al.} (\bibinfo {collaboration} {STAR}),\ }\href {\doibase
  10.1103/PhysRevLett.93.252301} {\bibfield  {journal} {\bibinfo  {journal}
  {Phys. Rev. Lett.}\ }\textbf {\bibinfo {volume} {93}},\ \bibinfo {pages}
  {252301} (\bibinfo {year} {2004})}\BibitemShut {NoStop}%
\bibitem [{\citenamefont {Cacciari}\ \emph {et~al.}(2005)\citenamefont
  {Cacciari}, \citenamefont {Nason},\ and\ \citenamefont {Vogt}}]{FONLL}%
  \BibitemOpen
  \bibfield  {author} {\bibinfo {author} {\bibfnamefont {M.}~\bibnamefont
  {Cacciari}}, \bibinfo {author} {\bibfnamefont {P.}~\bibnamefont {Nason}}, \
  and\ \bibinfo {author} {\bibfnamefont {R.}~\bibnamefont {Vogt}},\ }\href
  {http://www.lpthe.jussieu.fr/~cacciari/fonll/fonllform.html} {\bibfield
  {journal} {\bibinfo  {journal} {Phys. Rev. Lett.}\ }\textbf {\bibinfo
  {volume} {95}},\ \bibinfo {pages} {122001} (\bibinfo {year}
  {2005})}\BibitemShut {NoStop}%
\bibitem [{\citenamefont {Aggarwal}\ \emph {et~al.}(2010)\citenamefont
  {Aggarwal} \emph {et~al.}}]{STAReh}%
  \BibitemOpen
  \bibfield  {author} {\bibinfo {author} {\bibfnamefont {M.}~\bibnamefont
  {Aggarwal}} \emph {et~al.} (\bibinfo {collaboration} {STAR}),\ }\href
  {\doibase 10.1103/PhysRevLett.105.202301} {\bibfield  {journal} {\bibinfo
  {journal} {Phys. Rev. Lett.}\ }\textbf {\bibinfo {volume} {105}},\ \bibinfo
  {pages} {202301} (\bibinfo {year} {2010})}\BibitemShut {NoStop}%
\bibitem [{\citenamefont {Adare}\ \emph {et~al.}(2016)\citenamefont {Adare}
  \emph {et~al.}}]{PHENIXecb}%
  \BibitemOpen
  \bibfield  {author} {\bibinfo {author} {\bibfnamefont {A.}~\bibnamefont
  {Adare}} \emph {et~al.} (\bibinfo {collaboration} {PHENIX}),\ }\href
  {\doibase 10.1103/PhysRevC.93.034904} {\bibfield  {journal} {\bibinfo
  {journal} {Phys. Rev.}\ }\textbf {\bibinfo {volume} {C93}},\ \bibinfo {pages}
  {034904} (\bibinfo {year} {2016})}\BibitemShut {NoStop}%
\bibitem [{\citenamefont {Esha}\ \emph {et~al.}(2016)\citenamefont {Esha},
  \citenamefont {Nasim},\ and\ \citenamefont {Huang}}]{Nasim:2016}%
  \BibitemOpen
  \bibfield  {author} {\bibinfo {author} {\bibfnamefont {R.}~\bibnamefont
  {Esha}}, \bibinfo {author} {\bibfnamefont {M.}~\bibnamefont {Nasim}}, \ and\
  \bibinfo {author} {\bibfnamefont {H.}~\bibnamefont {Huang}},\ }\href@noop {}
  {\  (\bibinfo {year} {2016})},\ \Eprint {http://arxiv.org/abs/1603.02700}
  {arXiv:1603.02700 [nucl-th]} \BibitemShut {NoStop}%
\bibitem [{\citenamefont {Molnar}\ and\ \citenamefont
  {Voloshin}(2003)}]{NCQscaling}%
  \BibitemOpen
  \bibfield  {author} {\bibinfo {author} {\bibfnamefont {D.}~\bibnamefont
  {Molnar}}\ and\ \bibinfo {author} {\bibfnamefont {S.}~\bibnamefont
  {Voloshin}},\ }\href {\doibase 10.1103/PhysRevLett.91.092301} {\bibfield
  {journal} {\bibinfo  {journal} {Phys. Rev. Lett.}\ }\textbf {\bibinfo
  {volume} {91}},\ \bibinfo {pages} {092301} (\bibinfo {year}
  {2003})}\BibitemShut {NoStop}%
\bibitem [{\citenamefont {Afanasiev}\ \emph {et~al.}(2007)\citenamefont
  {Afanasiev} \emph {et~al.}}]{PHENIXmTscaling}%
  \BibitemOpen
  \bibfield  {author} {\bibinfo {author} {\bibfnamefont {S.}~\bibnamefont
  {Afanasiev}} \emph {et~al.} (\bibinfo {collaboration} {PHENIX}),\ }\href
  {\doibase 10.1103/PhysRevLett.99.052301} {\bibfield  {journal} {\bibinfo
  {journal} {Phys. Rev. Lett.}\ }\textbf {\bibinfo {volume} {99}},\ \bibinfo
  {pages} {052301} (\bibinfo {year} {2007})}\BibitemShut {NoStop}%
\bibitem [{\citenamefont {Pang}\ \emph {et~al.}(2015)\citenamefont {Pang},
  \citenamefont {Hatta}, \citenamefont {Wang},\ and\ \citenamefont
  {Xiao}}]{hydroPang}%
  \BibitemOpen
  \bibfield  {author} {\bibinfo {author} {\bibfnamefont {L.~G.}\ \bibnamefont
  {Pang}}, \bibinfo {author} {\bibfnamefont {Y.}~\bibnamefont {Hatta}},
  \bibinfo {author} {\bibfnamefont {X.~N.}\ \bibnamefont {Wang}}, \ and\
  \bibinfo {author} {\bibfnamefont {B.~W.}\ \bibnamefont {Xiao}},\ }\href
  {\doibase 10.1103/PhysRevD.91.074027} {\bibfield  {journal} {\bibinfo
  {journal} {Phys. Rev.}\ }\textbf {\bibinfo {volume} {D91}},\ \bibinfo {pages}
  {074027} (\bibinfo {year} {2015})}\BibitemShut {NoStop}%
\bibitem [{\citenamefont {Pang}\ \emph {et~al.}(2012)\citenamefont {Pang},
  \citenamefont {Wang},\ and\ \citenamefont {Wang}}]{hydroPang2}%
  \BibitemOpen
  \bibfield  {author} {\bibinfo {author} {\bibfnamefont {L.~G.}\ \bibnamefont
  {Pang}}, \bibinfo {author} {\bibfnamefont {Q.}~\bibnamefont {Wang}}, \ and\
  \bibinfo {author} {\bibfnamefont {X.~N.}\ \bibnamefont {Wang}},\ }\href
  {\doibase 10.1103/PhysRevC.86.024911} {\bibfield  {journal} {\bibinfo
  {journal} {Phys. Rev.}\ }\textbf {\bibinfo {volume} {C86}},\ \bibinfo {pages}
  {024911} (\bibinfo {year} {2012})}\BibitemShut {NoStop}%
\bibitem [{hyd()}]{hydroPrivateCom}%
  \BibitemOpen
  \href@noop {} {}\bibinfo {howpublished} {and private
  communication}\BibitemShut {NoStop}%
\bibitem [{\citenamefont {Banerjee}\ \emph {et~al.}(2012)\citenamefont
  {Banerjee}, \citenamefont {Datta}, \citenamefont {Gavai},\ and\ \citenamefont
  {Majumdar}}]{latticeBanerjee}%
  \BibitemOpen
  \bibfield  {author} {\bibinfo {author} {\bibfnamefont {D.}~\bibnamefont
  {Banerjee}}, \bibinfo {author} {\bibfnamefont {S.}~\bibnamefont {Datta}},
  \bibinfo {author} {\bibfnamefont {R.}~\bibnamefont {Gavai}}, \ and\ \bibinfo
  {author} {\bibfnamefont {P.}~\bibnamefont {Majumdar}},\ }\href {\doibase
  10.1103/PhysRevD.85.014510} {\bibfield  {journal} {\bibinfo  {journal} {Phys.
  Rev.}\ }\textbf {\bibinfo {volume} {D85}},\ \bibinfo {pages} {014510}
  (\bibinfo {year} {2012})}\BibitemShut {NoStop}%
\bibitem [{\citenamefont {Ding}\ \emph {et~al.}(2015)\citenamefont {Ding},
  \citenamefont {Karsch},\ and\ \citenamefont {Mukherjee}}]{latticeDing}%
  \BibitemOpen
  \bibfield  {author} {\bibinfo {author} {\bibfnamefont {H.~T.}\ \bibnamefont
  {Ding}}, \bibinfo {author} {\bibfnamefont {F.}~\bibnamefont {Karsch}}, \ and\
  \bibinfo {author} {\bibfnamefont {S.}~\bibnamefont {Mukherjee}},\ }\href
  {\doibase 10.1142/S0218301315300076} {\bibfield  {journal} {\bibinfo
  {journal} {Int. J. Mod. Phys.}\ }\textbf {\bibinfo {volume} {E24}},\ \bibinfo
  {pages} {1530007} (\bibinfo {year} {2015})}\BibitemShut {NoStop}%
\end{thebibliography}%

\end{document}